\documentclass[reprint,
superscriptaddress,
 amsmath,amssymb,
 aps,
 prl,
]{revtex4-2}

\usepackage[normalem]{ulem}

\newcommand\lya{$\mathrm{Ly}\alpha$}
\newcommand\mpc{h^{-1} \mathrm{Mpc}}
\newcommand\rpar{r_{\parallel}}
\newcommand\deltav{\delta_{v}}
\newcommand\biasd{b_{\delta}}

\usepackage{graphicx}
\usepackage{dcolumn}
\usepackage{hyperref}
\usepackage{subfigure}
\usepackage{xcolor}

\begin{document}

\title{First measurement of the correlation between cosmic voids and the Lyman-$\alpha$ forest}

\author{C. Ravoux}
\email{corentin.ravoux@clermont.in2p3.fr}
\affiliation{IRFU, CEA, Université Paris-Saclay, F-91191 Gif-sur-Yvette, France}
\affiliation{Université Clermont-Auvergne, CNRS, LPCA, 63000 Clermont-Ferrand, France}
\author{E. Armengaud}
\email{eric.armengaud@cea.fr}
\affiliation{IRFU, CEA, Université Paris-Saclay, F-91191 Gif-sur-Yvette, France}
\author{J. Bautista}
\affiliation{
 Aix Marseille Univ, CNRS/IN2P3, CPPM, Marseille, France}
\author{J. Rich}
\affiliation{IRFU, CEA, Université Paris-Saclay, F-91191 Gif-sur-Yvette, France}
\author{J.-M. Le Goff}
\affiliation{IRFU, CEA, Université Paris-Saclay, F-91191 Gif-sur-Yvette, France}
\author{N. Palanque-Delabrouille}
\affiliation{IRFU, CEA, Université Paris-Saclay, F-91191 Gif-sur-Yvette, France}
\affiliation{Physics Division, Lawrence Berkeley National Laboratory, Berkeley, CA 94720, USA}
\author{M. Walther}
\affiliation{University Observatory, Faculty of Physics,
Ludwig-Maximilians-Universität München, Scheinerstr. 1, 81679 Munich, Germany}
\author{C. Y\`eche}
\affiliation{IRFU, CEA, Université Paris-Saclay, F-91191 Gif-sur-Yvette, France}

\begin{abstract}
We report the first detection at a median redshift $z = 2.49$ of large-scale matter flows around cosmic voids. Voids are identified within a tomographic map of large-scale Lyman-$\alpha$ (\lya) transmissions, built from the eBOSS \lya~forest sample in the SDSS Stripe 82 field. We measure the imprint of flows around voids, known as redshift-space distortions (RSD), with a statistical significance of $8\,\sigma$. The observed quadrupole of the void-forest cross-correlation is described by a linear RSD model. The derived RSD parameter of the \lya~forest around voids is $\beta = 1.21 \pm 0.18$. Our model accounts for the tomographic effect induced by the \lya~data being located along parallel quasar lines of sight. This work presents a novel approach to observing the growth of cosmic structures at redshifts currently inaccessible to galaxy surveys.
\end{abstract}

\maketitle

\textit{Introduction} -- Large under-dense regions in the Universe, known as voids, provide an appealing way to study the cosmological evolution of large-scale structure. While structure formation is mostly studied using matter over-densities, e.g. from galaxy clustering measurements~\cite{DR162021}, voids were also used as a complementary probe, using in particular the galaxy-void cross-correlation at redshift $z<1.5$~\cite{Cai2016,Hamaus2017,Nadathur2019,Aubert2020,Hamaus:2020cbu,Euclid:2023eom}. At distances smaller than typical void sizes, the cross-correlation is proportional to the void density profile, while at mildly larger distances, redshift-space distortions (RSD) make the cross-correlation an indicator of the velocity flow around voids. Let $\deltav(r)$ be the mean matter density contrast at a distance $r$ from void centers. The associated average velocity flow in the linear regime reads~\cite{Hamaus2017}:

\begin{equation}
    \label{eq:velocity}
    {\bf v} = -\frac{1}{3} \frac{f H}{1+z}\overline{\deltav}(r){\bf r}\, ,
    \hspace*{5mm}
    \overline{\deltav}(r) = \frac{3}{r^3}\int_{0}^{r}dy y^2\deltav(y)\, ,
\end{equation}

\noindent where $H$ is the Hubble rate and $f$ is the linear growth rate of the large-scale structure. This flow generates a  quadrupole in the cross-correlation between void positions and the matter density field, whose amplitude is proportional to $f(\deltav(r)-\overline{\deltav}(r))$ \cite{Hamaus2017}. A key advantage of using voids is that, at a given scale, linear theory is a better description of velocity flows~\cite{Schuster:2022ogh} than around overdensities.

In this letter, we study for the first time RSD around voids at $z > 2$, i.e., at an earlier epoch than that studied with galaxy-void cross-correlations. We make use of the Lyman-$\alpha$ (\lya) forest observed in optical quasar spectra. This feature results from the absorption of light from background quasars by intervening neutral hydrogen and serves as a probe of large-scale structure. Large-scale 3D correlations in the \lya~forest already provided unique cosmological results at $z\sim 2-3$~\cite{duMasDesBourboux2020,DESIBAOlya2024,DESIBAOlya2025}. We provide here a new, different measurement which builds on a tomographic map of matter density and a derived void catalog computed from a sample of \lya~forests described in~\cite{Ravoux2020}.

\textit{Measurement} -- We use the \lya~forest region from 8199 quasar spectra from the SDSS-IV eBOSS survey's Data Release 16~ \cite{SDSS:2017yll,Dawson:2015wdb,DR162019}. To maximize the homogeneity and efficiency of void detection from this data, and to provide a statistically meaningful measurement, quasars in the so-called Stripe 82 field are used. In this 220~deg$^2$ field~\cite{Ravoux2020}, the total density of \lya\ lines of sight is $37\,{\rm{deg}^{-2}}$, about twice the eBOSS average.

\begin{figure*}[t]
    \centering
    \begin{subfigure}
        \centering
        \includegraphics[trim=9.5cm 0cm 0cm 0cm, clip=true,width = 0.42\textwidth]{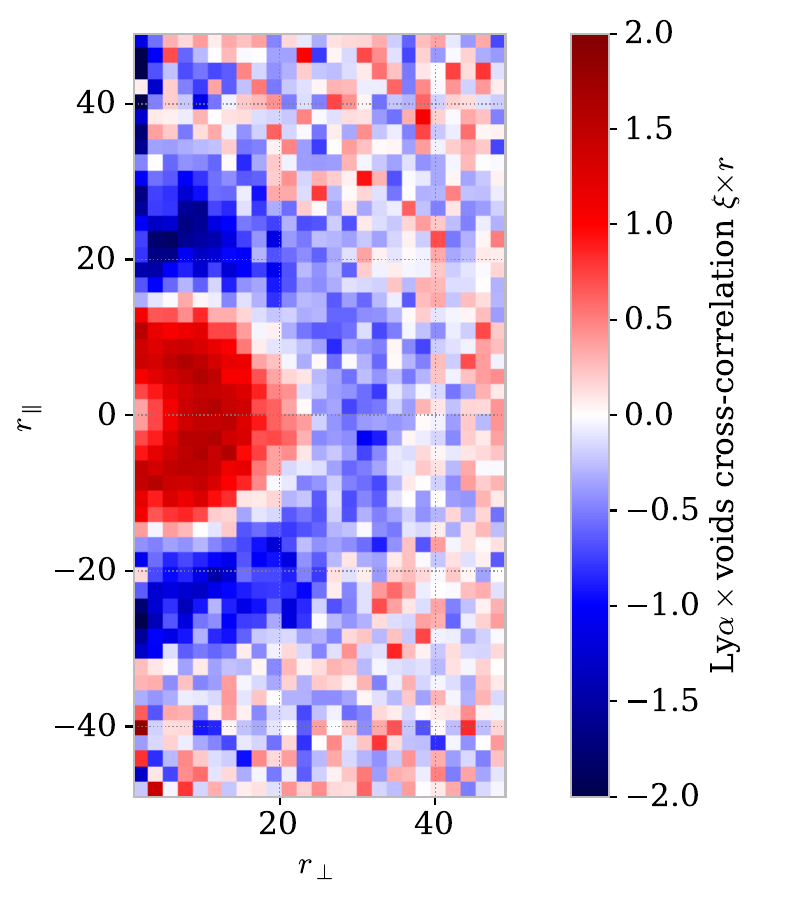}
    \end{subfigure}
    \begin{subfigure}
        \centering
        \includegraphics[trim=0cm 0cm 0cm 0cm, clip=true,width = 0.47\textwidth]{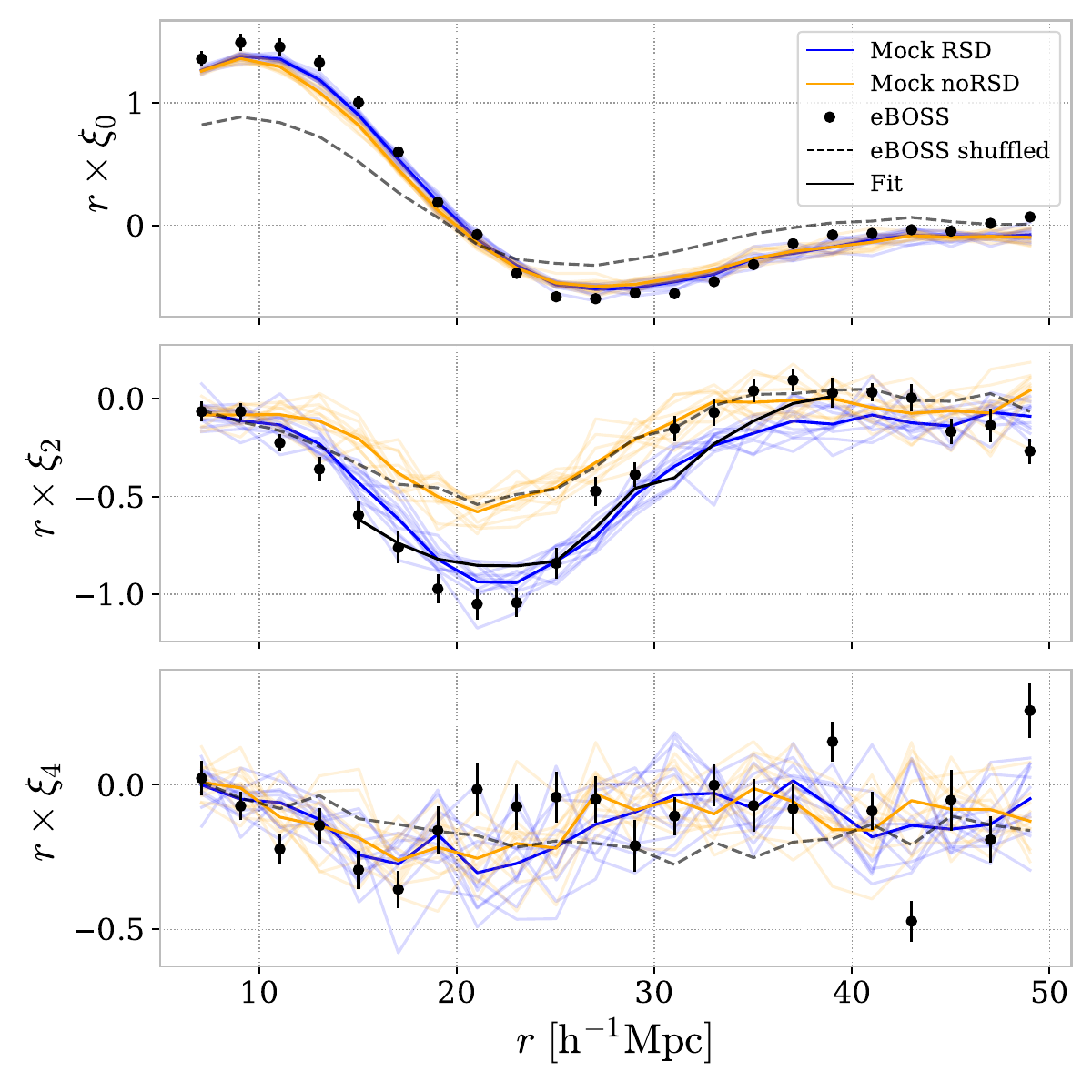}
    \end{subfigure}
    \caption{Left: measurement of $r\times \xi_(r_{\perp}, r_{\parallel})$ from eBOSS Stripe 82 data, where $r_\perp$ (resp. $r_\parallel$) is the transverse (resp. longitudinal) component of the separation vector with respect to the line of sight. Right: associated multipoles for $\ell=0, 2, 4$, from data (black points) and mock realizations including RSD (blue) or not (orange). Thin curves represent individual mock realizations, and their average is shown with thick curves. Black dashed curves show the average multipoles measured from shuffled eBOSS data. The black continuous curve shows the fit of the eBOSS quadrupole with Eqn.~\ref{eq:fit_model}.}
\label{fig:xcorr_void_lya}
\end{figure*}

Starting from quasar spectra, we estimate the \lya~flux contrast $\delta_{F}(\lambda) = F/\overline{F} -1$ as well as its variance $\sigma^2_{\delta_{F}}$, where $F(\lambda)$ is the \lya~transmission flux fraction along lines of sight, by following \cite{duMasDesBourboux2020}. From these, a 3D tomographic map of the \lya~forest flux contrast, $\delta_{F,{\rm tomo}}(x,y,z)$, is built in a grid of comoving coordinates. The algorithm uses a conjugate gradient Wiener filter to interpolate lines of sight with a Gaussian kernel weighted by $\sigma^{-2}_{\delta_{F}}$. The smoothing length of this kernel is $13\,\mpc$, chosen to match the mean angular separation between lines of sight. Coordinates are computed in the $\Lambda$CDM model with $\Omega_m = 0.3147$~\cite{Planck:2018vyg}. Void positions are found within our tomographic map using a simple spherical algorithm \cite{Stark2015a,Ravoux2020,Krolewski:2017jsm}: for each point where $\delta_{F,{\rm tomo}} > 0.14$, a spherical void is grown until the average flux contrast inside the volume reaches the value $0.12$, which defines the void radius. A void catalog is then created, keeping only the barycenters of the largest overlapping voids and by applying cuts detailed in~\cite{Ravoux2020}. We find 2906 voids located in the redshift range $z=2.1 - 3.2$, with a median redshift $z_{\rm med}=2.49$. Void radii are in the range $7 - 35\,\mpc$, with a median radius of $13.0\,\mpc$.

We measure the void-forest cross-correlation, $\xi(r,\mu)$, over the range $0<r<50~\mpc$ and $-1<\mu<1$, where $\mu=\rpar/r$ with $\rpar$ being the component along the line of sight of the separation of a forest pixel from a void center position. The bin widths are $\Delta r=2~\mpc$ and $\Delta\mu=0.04$. For a given bin $(r,\mu)$, the estimator \cite{font2012} is

\begin{equation}
 \xi(r,\mu) = \sum\limits_{(i,j)\in (r,\mu)} w_i \delta_{F}(\lambda_i) \; / \sum\limits_{(i,j)\in (r,\mu)} w_i\, ,
\end{equation}

\noindent where the sum is over voids (index $j$) and \lya-forest in quasar spectra (index $i$) at wavelength $\lambda_i$ such that the pair $(i,j)$ is contained in the $(r,\mu)$ bin.  The weights $w_i$ are chosen following \cite{duMasDesBourboux2020} to favor low-noise pixels and pixels at high redshift where the correlations are the strongest.
The covariance matrix of $\xi$ is determined using the subsampling method \cite{duMasDesBourboux2020} in which the Stripe 82 footprint is divided into non-overlapping sky regions. We find that this covariance estimator agrees to within 10\% with those derived from the variations of the $\xi$ in 11 mock realizations described below.

To quantify the angular asymmetry of $\xi(r,\mu)$, it is convenient to compute its multipole expansion onto the Legendre polynomial basis, $\xi_\ell(r)$, specifically the monopole $\xi_0$, quadrupole $\xi_2$, and hexadecapole $\xi_4$. Fig.~\ref{fig:xcorr_void_lya} shows the measured 2D correlation as well as its multipoles. The positive monopole for $r\lesssim20\mpc$ reflects the typical size of the voids, while the negative values for $r>20\mpc$ reflect the overdensities outside the voids. A strong quadrupole is measured at separation $r\sim20\mpc$.

\textit{Simulations} -- The correlations of the eBOSS data in  Fig.~\ref{fig:xcorr_void_lya} cannot be directly compared to a physical model because they are contaminated by several measurement artifacts, primarily due to the small density of lines-of-sight compared to the high density of flux  measurements within individual lines-of-sight. Interpretation of the data therefore depends on the use of synthetic \lya~samples ("Saclay mocks")~\cite{Etourneau2023}. Those contain lines of sight with correlated \lya~transmission flux fractions
$F(\lambda)$.
The $F(\lambda)$ are generated using two Gaussian-random fields: $\delta_L$ generated on a coarse-grained 3D grid ($2.19 \mpc$ cell size), and $\delta_S$ on a fine-grained 1D grid ($0.2 \mpc$ cell size). Along quasar lines of sight, $F(\lambda)$ is computed from $\delta_L(\lambda)$ and $\delta_S(\lambda)$~\cite{weinberg_lower_1997}:

\begin{equation}
     F(\lambda) = \mathrm{exp}\left[ - a_{\mathrm{GP}}\, \mathrm{exp}\left(b_{\mathrm{GP}} (\delta_{L} + \delta_{S} + c_{\mathrm{GP}}\eta_{\parallel})\right)\right]\, ,
    \label{eq:fgpa}
\end{equation}

\noindent where $\eta_{\parallel}$ is the velocity gradient associated with $\delta_L$ along the line of sight, in the linear approximation. The parameters $a_{\mathrm{GP}}(z)$, $b_{\mathrm{GP}}(z)$ and $c_{\mathrm{GP}}(z)$ are empirical coefficients tuned to reproduce the observed statistical properties of the \lya~forest: the mean transmitted flux fraction $\overline{F}$, the large-scale \lya~bias and its RSD parameter as measured in~\cite{duMasDesBourboux2020}.

Mock spectra using Eqn.~\ref{eq:fgpa} with $\delta_F(\lambda)=F(\lambda)/\overline{F}-1$ constitute the first level of mocks, hereafter called ``raw'' mocks. Those neglect several astrophysical and instrumental effects that must be incorporated to make realistic mocks. First, for real data, the estimation of $\delta_F(\lambda)$ relies on a model for the background quasar intrinsic continuum spectrum. The latter biases the flux contrast estimate, specifically for the method we use~\cite{duMasDesBourboux2020,Busca2025}, which leads to a characteristic ``continuum-fitting distortion" of correlations. From the "raw" mocks, we therefore multiply their $F(\lambda)$ by a quasar continuum model, and $\delta_F(\lambda)$ is then estimated using the same continuum fitting pipeline as for the data. Second, we add instrumental noise to the spectra. Finally, we add strong absorbers (HCD objects) to the spectra, following~\cite{Chabanier2021, Font-Ribera:2012qgn} and correlated absorption by metals following eg.~\cite{Herrera-Alcantar:2023its}.

We generated eleven mock realizations of the full Stripe 82 dataset, labeled RSD-mocks, including all above-mentioned contaminants. Eleven additional realizations, labeled noRSD-mocks, were produced with the same method but setting $c_{\mathrm{GP}} = 0$, i.e., without the effect of the large-scale velocity flow. For each mock realization, reconstructed flux contrasts $\delta_F(\lambda)$, void catalogs, and $\xi(r,\mu)$ are computed in the same way as for the eBOSS data. The individual realizations and average values of mock multipoles are shown on Fig.~\ref{fig:xcorr_void_lya}, both for RSD- and noRSD-mocks. The addition of the RSD effect impacts essentially $\xi_2$: we observe a negative quadrupole in noRSD-mocks, and a different, significantly larger $\xi_2$ in absolute value for the RSD-mocks. As we will see below, the quadrupole in the noRSD-mocks is primarily generated by the \lya~line of sight geometry, while that in the RSD-mocks is also impacted by the velocity flow. On the other hand the monopole is only mildly affected by the RSD. A hexadecapole signal is found on the average mocks, but its amplitude is similar to individual mocks fluctuations. Finally, we observe that the data multipoles are in reasonable agreement with the RSD mocks.

\textit{Modeling $\xi$} -- The measured anisotropies of the void-forest cross-correlation in redshift space are due to cosmological peculiar velocities (RSD), but also to the sparse transverse sampling of the \lya~forest signal due to the limited number of quasar lines of sight. For the RSD effect, we use the same approach as eg.~\cite{Cai2016} for the void-galaxy correlation (see Appendix A in the Supplemental material). We model the void-forest cross-correlation in redshift space as

\begin{equation}
    \xi(r,\mu) = \biasd \deltav(r) + b_{\eta}f \left[ \frac{ \overline{\deltav}(r)}{3} + 
    \left(\deltav(r) - \overline{\deltav}(r)\right) \mu^2 \right]\, ,
\label{eq:xi}
\end{equation}

\noindent where $\biasd$ and $b_\eta$ are unknown biases relating $\delta_F$ to fluctuations of density and to the longitudinal velocity gradient. Defining the RSD term $\beta = b_\eta f/\biasd$, and $\overline{\xi_n}(r) = \frac{3}{r^3} \int_0^r dy y^2 \xi_n(y)$, it follows that:

\begin{equation}
    \xi_2(r) = \frac{2\beta}{3+\beta} \left(\xi_0(r) - \overline{\xi_0}(r) \right).
\label{eq:xi_2}
\end{equation}

Unfortunately, the geometry of the \lya~forest sample and the details of the void finder also impact the quadrupole $\xi_2(r)$. Sparse sampling of forests in the direction transverse to the line of sight results in a systematic displacement of the reconstructed void center positions towards the nearest line of sight, relative to their true positions. Additionally, the precision of position reconstruction is anisotropic with respect to the lines of sight. In Appendix B of the Supplemental material, we present simple models to study these effects, labeled tomographic. They have a geometrical origin and are therefore taken into account in mock realizations: in particular they are the source of the ``tomographic'' quadrupole seen in the noRSD-mocks, in Fig.~\ref{fig:xcorr_void_lya}. To test for the sparse line of sight origin of this quadrupole, we used a simple Gaussian mock realization with a very dense grid of lines of sight as an input to the void finder. We found that the maximum value for the quadrupole-to-monopole ratio is 3~\% in that case, which is more than an order of magnitude smaller than that seen in the noRSD-mocks.

\begin{figure}[!t]
    \centering
    \includegraphics[trim=0.5cm 0.5cm 0cm 0cm, clip=true,width = 0.48\textwidth]{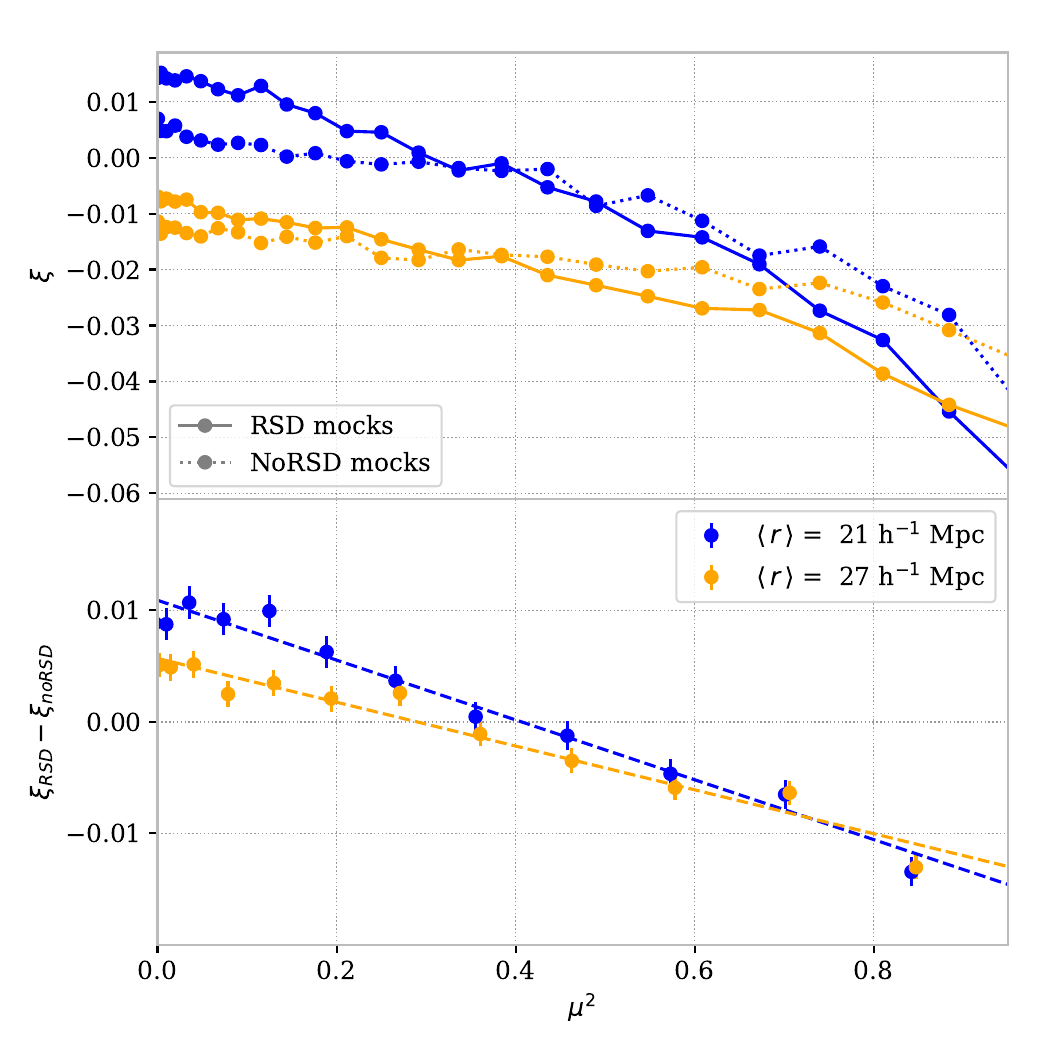}
    \caption{The cross-correlation $\xi$ as a function of $\mu^2$ for mock realizations, for the  bins $20< r<22$ and $26<r<28~\mpc$. Top: results for RSD-mocks (continuous, $\xi_{\rm RSD}$) and noRSD-mocks (dotted, $\xi_{\rm noRSD}$). Bottom: difference $\xi_{\rm RSD}-\xi_{\rm noRSD}$, with statistical error bars. Dashed lines show associated linear fits.}
\label{fig:kaiser_plot}
\end{figure}

To illustrate how RSD can be disentangled from the tomographic effect, Fig.~\ref{fig:kaiser_plot} (top) shows $\xi(r,\mu)$ as a function of $\mu^2$ obtained from mocks, with and without RSD. Following Eqn.~\ref{eq:xi}, in the absence of the tomographic effect, $\xi$ would be independent of $\mu$ in the noRSD case, and a linear function of $\mu^2$ in the RSD case. This picture is too simple due to the tomographic effect, as can be seen in the top panel of Fig.~\ref{fig:kaiser_plot}. On the other hand, Fig.~\ref{fig:kaiser_plot} (bottom) shows that the difference $\xi_{\rm RSD} - \xi_{\rm noRSD}$ is approximately a linear function of $\mu^2$: this suggests that the tomographic effect is an additive contribution to the quadrupole.

Finally, we also performed a crucial ``shuffling'' test on both mocks and data where individual lines of sight were randomly permuted within redshift bins of width $\Delta z=0.1$. The shuffling preserves the geometry and noise properties of the sample while eliminating all physical void-forest and forest-forest correlations. A void catalog and associated $\xi$ were computed from these shuffled $\delta_F$. The dashed lines on Fig.~\ref{fig:xcorr_void_lya} show the multipoles from shuffled data. As expected, the shuffling reduces the magnitude of both the monopole and the quadrupole. In the case of mocks, we observe that the quadrupole measured after shuffling the RSD-mocks is close to that measured from the unshuffled noRSD-mocks. Therefore, we conclude that the quadrupole from shuffled correlations can be considered an approximation of the quadrupole induced by the tomographic effect.

The average hexadecapole measured from the mocks is the same within statistical uncertainties, whether or not the mocks include RSD: we do not attribute this hexadecapole to the velocity flow, which is consistent with the linear flow model prediction $\xi_4 = 0$ (Eqn.~\ref{eq:xi}).
The hexadecapole in eBOSS data is consistent with mocks, but given the statistical uncertainties, it cannot be distinguished from zero for $r \gtrsim 20\,\mpc$.

\textit{Measuring the RSD effect from data} -- We now present a simple model fit to the measured quadrupole, using a similar approach to that used for galaxies, e.g., in~\cite{Cai2016,Hamaus2017}: we estimate the \lya~RSD parameter $\beta$ from Eqn.~\ref{eq:xi_2}, taking the tomographic effect into account.
As suggested in previous paragraphs, we decide here to rely on the shuffling method to estimate the impact of this effect on the multipoles: both the monopole and quadrupole of the shuffle data, $\xi_{0,s}(r)$ and $\xi_{2,s}(r)$ (dashed lines in Fig.~\ref{fig:xcorr_void_lya}), are assumed to be additive corrections to the multipoles in Eqn.~\ref{eq:xi_2}.

After including those empirical data-driven corrections, we therefore get the following relation:

\begin{align}
    \xi_2(r) & = \xi_{2,s}(r) + \frac{2\beta}{3+\beta} \nonumber \\
    & \times \left[ (\xi_0(r) - \xi_{0,s}(r)) - \overline{(\xi_0(r) - \xi_{0,s}(r))}\, \right]\, .
    \label{eq:fit_model}
\end{align}

The RSD term $\beta = b_{\eta}f/b_{\delta}$ is a single free parameter. The fit is done for $15 \leq r \leq 40\,\mpc$, i.e. we exclude the region $r<15\,\mpc$ determined by the inner void profile, and we use statistical errors from the covariance matrix of $\xi_2$.

We first test this procedure on the average RSD-mock correlations, and find a good fit ($\chi^2$ p-value of 0.3) with $\beta = 1.41\pm 0.16$. The related RSD parameter, as measured from the \lya~auto-correlation using similar mock data in~\cite{Etourneau2023}, was found to be $\beta \simeq 1.5$ at redshift $z \simeq 2.5$. This indicates that our estimate for $\beta$ is unbiased within $\sim 1\sigma$ statistical uncertainties, therefore validating the empirical shuffling-based correction.

Turning to the data, with the same method we derive $\beta = 1.08 \pm 0.13$. 
This corresponds to a detection of the RSD effect formally at $8\,\sigma$ statistical significance. The fit is shown as a continuous black line on Fig.~\ref{fig:xcorr_void_lya}. The model for the quadrupole is not a very good match to the data ($\chi^2$ p-value of $3\times10^{-3}$), although it is a good match to the average quadrupole of RSD-mocks.
As a variation, we replace $\xi_{2,s}$ in Eqn~\ref{eq:fit_model} with the average quadrupole of the noRSD-mocks, which has less statistical fluctuations: the measured value remains statistically consistent, $\beta = 1.16\pm 0.14$.

\begin{figure}
    \centering
    \includegraphics[trim=0cm 0cm 0cm 0cm, clip=true,width = 0.48\textwidth]{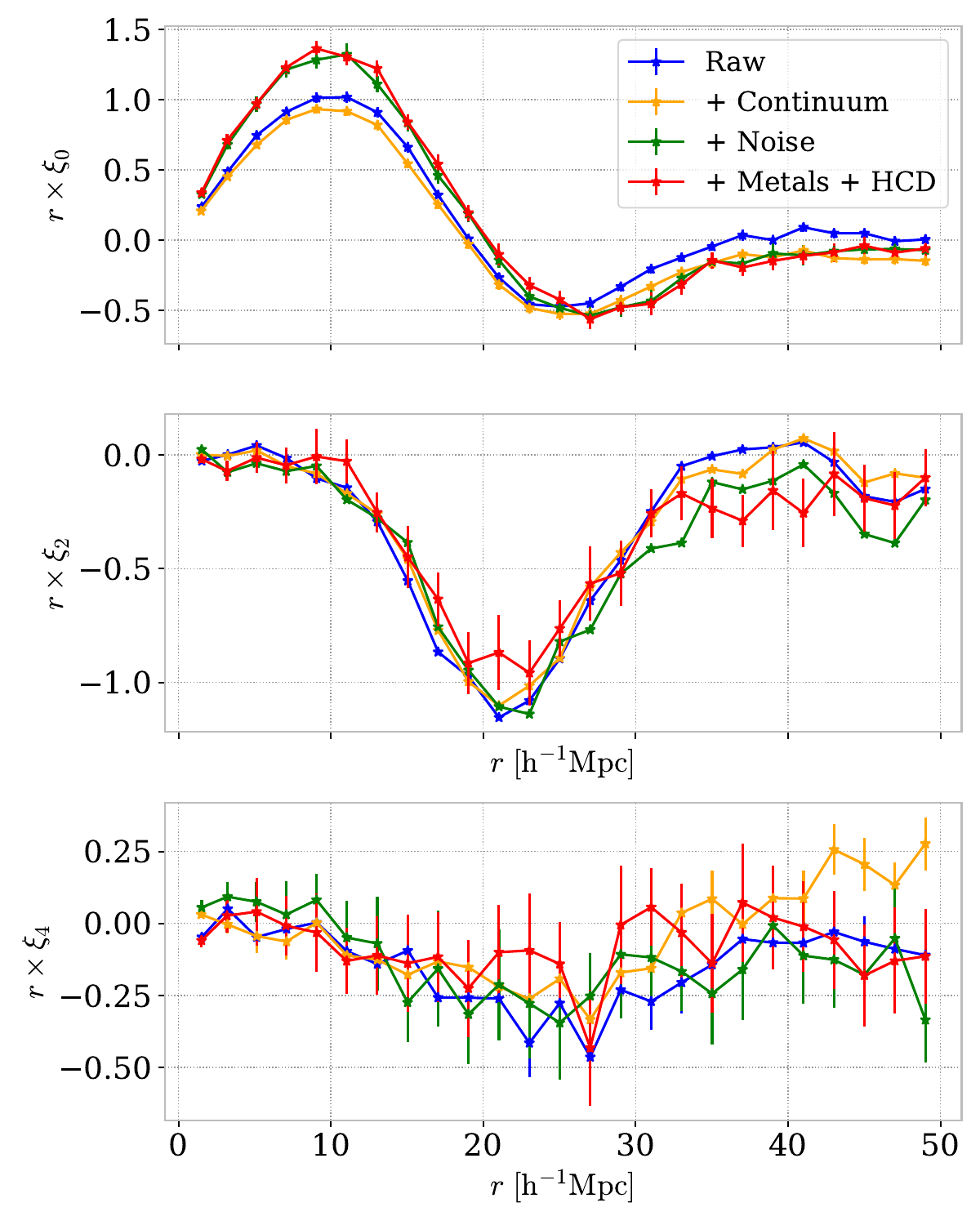}
    \caption{Measured monopoles, quadrupoles, and hexadecapoles of $\xi$ in mock data, including different instrumental and astrophysical effects. "Raw" mocks (blue) include only the absorption law from Eqn.~\ref{eq:fgpa}. Statistical error bars are estimated using sub-sample variance. For the quadrupole, they are only represented on the red curve.}
\label{fig:systematic_study_quadrupole}
\end{figure}

\textit{Other systematics} --
As illustrated by Fig.~10 in~\cite{Ravoux2020}, voids with radii $\lesssim 15\,\mpc$ are contaminated by noise fluctuations. Using mock data, we performed the same study after selecting voids with radius $>15\,\mpc$, keeping 40\% of the void statistics.
With respect to the baseline measurement, the radial dependence of multipoles is scaled because of the larger average void radius, e.g. the separation at which $\xi_0 = 0$ is shifted from 20.4 to $24.3 \,\mpc$. However, the statistical error on the multipoles is increased by 60\%, which justifies the choice to include smaller voids for this specific statistics-limited study. Also, we expect the monopole correction with shuffling to mitigate this effect, since $\xi_{0,s}$ is identically contaminated by the noise.

We now turn to other systematic effects that are specific to the \lya~forest, and were briefly mentioned when we described our mock sample.
The multipoles of $\xi$ are shown in Fig.~\ref{fig:systematic_study_quadrupole} for several types of mock realizations, from ``raw'' to ``fully contaminated''. Their main features remain unchanged. 
Including the effect of quasar continuum-fitting doesn't change $\xi_2$ within our statistical errors, but it increases $|\xi_0(r)|$ at large $r$, e.g. by a factor 30\% for $r=30\,\mpc$. We compared RSD fit results on this single mock realization with and without continuum fitting: we found that due to this change in $\xi_0$, the fit value of $\beta$ is decreased by 12~\% when including continuum fitting. Therefore, we choose to apply this 12~\% correction to our measured $\beta$, and we add in quadrature a 12~\% systematic uncertainty to the statistical error.
When noise is included, additional small-scale fluctuations in $\delta_F$ increase the number of small, spurious voids identified by the void finder. This results in an increase of $|\xi_0(r)|$ at small radius.

Finally, metals and HCD systems only marginally modify the cross-correlation. This later feature is an interesting property of the void-forest correlation.

\textit{Discussion and perspectives} -- From the Stripe 82 subsample of eBOSS data, we have measured the void-\lya~forest cross-correlation at a median redshift $z=2.49$. The measurement matches the prediction from simulated data, and we presented a simple model including the RSD effect in the linear regime. This is the first time that voids detected from the \lya~forest are used in the context of a large cosmological survey, and the first observation of large-scale velocity flows away from voids at such a high redshift. The effect is detected with an $\sim 8\,\sigma$ statistical significance. Remarkably, the linear model matches reasonably well mocks and data down to separation distances $r\sim 20\,\mpc$. This has been observed in simulations~\cite{Stark2015a} and highlights the advantage of using both voids and the high-redshift \lya~forest to study the linear growth of structure.

Taking into account the above-mentioned correction related to the
continuum fitting, the measured RSD-related amplitude is $\beta = 1.21 \pm 0.18$. This value is smaller than a similar parameter derived from the large-scale eBOSS \lya~auto-correlation, $\beta_{\rm auto} = 1.66 \pm 0.07$~\cite{duMasDesBourboux2020}. Comparisons are however difficult, eg. because of possible redshift evolution, and more recent measurements of $\beta_{\rm auto}$ from DESI yield smaller values~\cite{DESIBAOlya2025, Cuceu:2025nvl}.
We used the simplest possible RSD model to interpret the data, and found that it provides a reasonable description of mocks and data. However, we do not expect the mitigation of tomographic effects with the shuffling method to be perfect. One approach to avoid them in the future would be to use cross-correlations, i.e. between our \lya~forest voids and galaxy positions that are independent of the void finder. In addition, more realistic models, such as~\cite{Nadathur2019, Chuang:2016wqb}, could be used and may impact our measurement.

This exploratory work, applied to a statistically limited dataset, opens up new possibilities for observational cosmology. On the one hand, ongoing and future dense \lya~surveys~\cite{Japelj:2019ksp,Lee2018,Newman:2020iao} will more precisely probe the dynamics of small under-dense regions in the high-redshift cosmic web. On the other hand, ongoing large-field surveys such as WEAVE-QSO~\cite{WEAVE:2016rxg} and DESI~\cite{DESI:2016fyo} will extend this measurement to larger volumes. In addition to measurements of the linear cosmic flow at small scales and high redshifts, this opens the way to perform Alcock-Paczynsky tests with voids~\cite{Nadathur2019} at high redshift, complementary to~\cite{Cuceu:2025nvl}.

\begin{acknowledgments}
We thank Sesh Nadathur and Arnaud de Mattia for fruitful discussions. We acknowledge the use of the \texttt{picca}\footnote{\url{https://github.com/igmhub/picca}} and \texttt{quickquasars}\footnote{\url{https://github.com/desihub/desisim/blob/main/py/desisim/scripts/quickquasars.py}} softwares. The authors acknowledge support from grants ANR-16-CE31-0021 and ANR-22-CE92-0037. The project leading to this publication has received funding from Excellence Initiative of Aix-Marseille University - A*MIDEX, a French ``Investissements d'Avenir'' programme (AMX-20-CE-02 - DARKUNI). This work made use of granted access to the HPC resources of TGCC under the allocation A0070410586 made by GENCI (Grand Equipement National de Calcul Intensif).

\end{acknowledgments}

\bibliographystyle{apsrev4-1}
\bibliography{lyavoids_biblio}

\appendix

\subsection{Appendix A - Modelling the effect of RSD on $\xi$}

\textit{Notation} - Given a comoving separation vector ${\bf r}$ between a void center and another point, we note $(\rpar, {\bf r}_{\perp})$ its parallel and perpendicular projections with respect to the observer line of sight. The associated polar coordinates are $(r, \mu)$ where $r^2 =  \rpar^2 +{\bf r}_{\perp}^2$ and $\mu = \rpar/r$. For a function $f({\bf r}) = f(r,\mu)$, its multipole expansion onto the Legendre polynomials $P_{\ell}(\mu)$ is defined by:

$$f_{\ell}(r) = (2\ell+1) \int_0^1 f(r,\mu) P_{\ell}(\mu) d\mu$$

\noindent In particular, the non-vanishing coefficients for $f = \alpha + \beta \mu^2$ are $f_0 = \alpha+\beta/3$ and $f_2 = 2\beta/3$. Finally, for any function $f(x)$, we define its spherical volume average:
$$\overline{f}(x) = \frac{3}{x^3}\int_{0}^{x} t^2\, f(t)\, dt$$

\textit{Modelling} - In order to model the effect of large-scale cosmic flows on the void - \lya\, cross-correlation $\xi$, we adapt the simple approach from~\cite{Cai2016, Hamaus2017} to the case of the \lya~forest. This approach relies on a set of assumptions that are briefly presented here. Quantifying the validity of these assumptions goes beyond the scope of this article, as it requires dedicated cosmological simulations, including in some cases hydrodynamical simulations to model the \lya~forest.

First, we assume that the measured \lya~flux contrast $\delta_F$ is linearly related to fluctuations of the matter density $\delta_{\rm matter}$, and to the projection along the line of sight of the matter velocity field gradient, noted $\eta_{\parallel}$:

\begin{equation}
    \label{eq:linear_scaling}
\delta_F = b_{\delta} \delta_{\rm matter} + b_{\eta} \eta_{\parallel} + \delta_N
\end{equation}

\noindent In this expression, $\delta_N$ includes any correction to this linear scaling, in particular small-scale fluctuations in the \lya~forest which we assume are not correlated to the large-scale fields. The coefficients $b_{\delta}$ and $b_{\eta}$ are linear bias coefficients. We now consider the properties of the matter field around voids. In real space the matter density is isotropic on average, and we note $\langle \delta_{\rm matter}\rangle = \deltav(r)$ this average void profile as a function of $r$, the radial coordinate with respect to the void center. We then assume that the velocity field around voids follows the linear regime of structure formation. This implies that the average velocity field is~\cite{Peebles:1994xt}:

\begin{equation}
    \label{eq:velocity}
    \langle {\bf v} \rangle(r) = -\frac{1}{3} \frac{f H(z)}{1+z}\overline{\deltav}(r){\bf r}
\end{equation}

Although we consider here relatively small separation distances down to $r\sim 15 \,\mpc$, we expect this relation to be a fairly good approximation. Indeed, the dynamics within and around voids is more linear than around halos, and we consider only redshifts $z>2$, where large-scale formation is more linear than for low redshifts. The corresponding average parallel velocity gradient is then

\begin{equation}
    \langle \eta_{\parallel} \rangle ({\bf r}) \equiv \frac{-(1+z)}{H(z)} \frac{\partial v_{\parallel}}{\partial \rpar} = \frac{f \overline{\deltav}}{3} + f \mu^2 (\deltav - \overline{\deltav}) 
\end{equation}

Turning to the measured void - \lya~cross-correlation $\xi$, here we assume that the positions of void centers are perfectly reconstructed, so that, averaging over a large number of voids we have $\xi({\bf r}) = \langle \delta_F \rangle({\bf r})$. This results in
\begin{equation}
    \label{eq:xi_rsd}
    \xi(r,\mu) = b_{\delta} \deltav(r) + b_{\eta} f \left[ \frac{ \overline{\deltav}(r)}{3} + 
    \left(\deltav(r) - \overline{\deltav}(r)\right) \mu^2 \right]
\end{equation}

\noindent This is Eqn.~4 in the main article. The non-vanishing coefficients in the multipole expansion of $\xi$ are then $\xi_0 = (b_\delta +b_\eta f/3)\deltav$ and $\xi_2 = 2b_\eta f(\deltav-\overline{\deltav})/3$. As a consequence, one can directly express the quadrupole as a function of the monopole, recovering Eqn.~5 of the main article:

\begin{equation}
    \label{eq:xi2vsxi0}
    \xi_2(r) = \frac{2\beta}{3+\beta} \left(\xi_0(r) - \overline{\xi_0(r)}\right)  \;\;\;\;\mbox{where}\;\; \beta = b_\eta f/b_\delta
\end{equation}

This relation is formally identical to the one derived in the case of galaxy-void correlations, under similar assumptions~\cite{Hamaus2017}. However, in our case the $\beta$ coefficient depends on both bias parameters $b_\delta$ and $b_\eta$, so that the linear growth rate $f$ cannot be estimated from measurements of $\beta$ alone. The same happens when interpreting the \lya~auto-correlation function.
In the framework of this simple model, there is no RSD-induced hexadecapole in $\xi$ (which would be induced for example by the void center's motion).

\subsection{Appendix B - The effects of void reconstruction on $\xi$}

The positions of \lya~pixels may be considered as perfectly determined, since they are derived from measured angular positions and spectroscopic wavelengths. This is not the case of the void centers positions, for which the uncertainties are not negligible. Here, we present simple toy models to quantify these effects and justify the mitigation strategy used to measure $\beta$ in the main article.

\subsubsection{Resolution of the reconstructed void position}
The void-finding algorithm, with its anisotropic precision, has an impact on the void-forest correlations. As described in~\cite{Ravoux2020}, the tomographic map is not isotropic due to the geometry of parallel lines of sight. In this work we had found that a stack of the mock tomographic map, without RSD effects, around under-dense regions, had an axis ratio of 1.33. The radial coordinate of a void center is therefore better reconstructed than its transverse position. Let us assume that the resolution function of void positions is an asymmetric Gaussian of widths $(\sigma_\parallel, \sigma_\perp)$ with $\sigma_\perp \simeq 1.3 \,\sigma_\parallel$ in our case. Under this assumption, the measured $\xi$ relates to the unbiased ``raw'' correlation $\xi_{\rm raw}$ following

\begin{equation}
    \label{eq:smooth_aniso}
    \xi(r_\parallel, r_\perp) = \frac{1}{2\pi \sigma_\parallel \sigma_\perp} \int d^2{\bf x} \exp\left(-\frac{x_\parallel^2}{2\sigma_\parallel}-\frac{x_\perp^2}{2\sigma_\perp}\right) \xi_{\rm raw}({\bf r}-{\bf x})
\end{equation}

This anisotropic smoothing modifies both the monopole $\xi_0$ and quadrupole $\xi_2$, and introduces a non-zero quadrupole even if $\xi_{2,{\rm raw}}=0$. As a consequence, Eqn.~\ref{eq:xi2vsxi0} no longer holds. However, as explained in the article, before fitting Eqn.~\ref{eq:xi2vsxi0} the measured multipoles are subtracted from their shuffled values, which are computed after permutations of the input line of sight data. The shuffling procedure keeps the effect of anisotropic resolution, so that we expect it mitigates this effect. In order to illustrate this, we introduce a simple model, in which $\xi_{\rm raw}$ is constructed from the universal void profile of~\cite{Hamaus:2014fma}:

\begin{equation}
    \label{eq:voidprofile}
    \deltav(r) = \delta_c  \,\frac{1-(r/r_s)^{\alpha_v}}{1+(r/r_v)^{\beta_v}}
\end{equation}

\noindent The following parameters provide a good fit to the shape of the average monopole as derived from the eleven mock samples described in the article: $r_v=22\,\mpc$, $r_s = 0.82\,r_v$, $\alpha_v=0.7$, $\beta_v=7.8$, $\sigma_\parallel=3.9\,\mpc$ and $\sigma_\perp=5.1\,\mpc$.
Having this model for $\deltav$, we numerically compute $\xi_{\rm raw}(r,\mu)$ from Eqn.~\ref{eq:xi_rsd}, with the RSD parameter $\beta=1$. We then numerically derive the smoothed $\xi(r,\mu)$ and its multipoles $\xi_0$ and $\xi_2$, using Eqn.~\ref{eq:smooth_aniso}. Finally, we estimate the RSD parameter $\beta_{\rm meas}$ from the ratio $\xi_2/(\xi_0-\overline{\xi_0})$, using Eqn.~\ref{eq:xi2vsxi0}. Taking the median value in the range $15\leq r\leq 30\,\mpc$, we obtain $\beta_{\rm meas} = 1.25$: in this toy model, the anisotropic resolution increases the RSD estimator by roughly 25~\%.

On the other hand, we emulate the shuffling correction  with the following: we consider the same model as before, but without RSD effect, $\xi_{\rm noRSD}(r,\mu)$, by setting $\beta=0$. Its quadrupole $\xi_{2, \rm noRSD}(r)$ is subtracted to effectively correct for tomographic effects. Indeed, as noticed in the main article we observed that the \textit{shuffled} quadrupole for the RSD-mocks is close to the \textit{unshuffled} quadrupole of the noRSD-mocks. From the computed ratio $(\xi_2-\xi_{2, \rm noRSD})/(\xi_0-\overline{\xi_0})$, we derive the updated value $\beta_{\rm meas} = 1.03$: the additive "shuffling" correction to the quadrupole corrected for a large fraction of the initial bias.

\subsubsection{Accuracy of the reconstructed void position}
Given the specific line of sight geometry of the \lya~data, we expect that void positions inferred from any void finding algorithm are also biased with respect to the positions of the lines of sight. Such a bias was already highlighted in the case of galaxies~\cite{Massara:2022lng}. This effect differs from the anisotropic smoothing described above in Eqn.~\ref{eq:smooth_aniso}, in that it is a systematic bias and not a resolution effect.

In our case, voids are identified from a tomographic map created by Wiener filtering the input sample. We have observed that the average flux contrast of this map is smaller at locations further away from lines of sight, where less data is available to reconstruct the \lya~absorption signal. As a consequence, the reconstructed positions of void centers are found on average closer towards the nearest line of sight with respect to their true positions. As a toy-model, let us assume that the transverse distance of a reconstructed void to a given line of sight is reduced on average by an offset $D$ with respect to its true position. From our void catalog, we actually estimated $D\simeq 2.5\,\mpc$, using the measured distribution of distances between void centers and the nearest line of sight. As in previous subsection, we note $\xi_{\rm raw}$ the unbiased correlation, so that the measured correlation is given by

\begin{equation}\label{eq:voidbias}
    \xi(\rpar, r_{\perp}) = \xi_{\rm raw}(r_{\parallel}, r_{\perp}+D)    
\end{equation}

\noindent It is easy to see that this offset introduces a quadrupole even if $\xi_{\rm raw}$ is isotropic. For example, assuming that $D$ is small with respect to $r_\perp$, and that the quadrupole is small with respect to the monopole of $\xi_{\rm raw}$, we could expand this expression to first order:

\begin{equation}
\xi(r,\mu) \simeq \xi_{\rm raw}(r,\mu) + D\,\frac{d\xi_{0,{\rm raw}}}{dr}\sqrt{1-\mu^2}
\end{equation}

As in the case of an asymmetric resolution function, Eqn.~\ref{eq:xi2vsxi0} doesn't hold anymore, but we expect the shuffling procedure to mitigate this effect. To illustrate this, let us use the same procedure as above. We start from $\xi_{\rm raw}$, which we compute from Eqns.~\ref{eq:voidprofile} and~\ref{eq:xi_rsd}, with an input RSD parameter $\beta=1$. To model the bias of reconstructed void positions, we use Eqn.~\ref{eq:voidbias} with $D=2.5\,\mpc$, and then numerically derive the "measured" multipoles $\xi_0$ and $\xi_2$. Finally, we compute an estimated RSD parameter $\beta_{\rm meas}$, from the ratio $\xi_2/(\xi_0-\overline{\xi_0})$ (Eqn.~\ref{eq:xi2vsxi0}). Using the same procedure as before, without any correction we obtain $\beta_{\rm meas}=1.11$. In this toy model, the bias on reconstructed void positions increases the RSD estimator by 11~\%.

Then, as before we emulate the shuffling correction by making use of the quadrupole computed from the same toy model, except it has $\beta=0$. After this correction we obtain $\beta_{\rm meas}=1.03$: again, a large fraction of the bias was corrected thanks to this procedure.

\subsection{Appendix C - Qualitative comparison with the void-galaxy correlation}

\begin{figure}
    \centering
    \includegraphics[trim=0cm 0cm 0cm 0cm, clip=true,width = 0.5\textwidth]{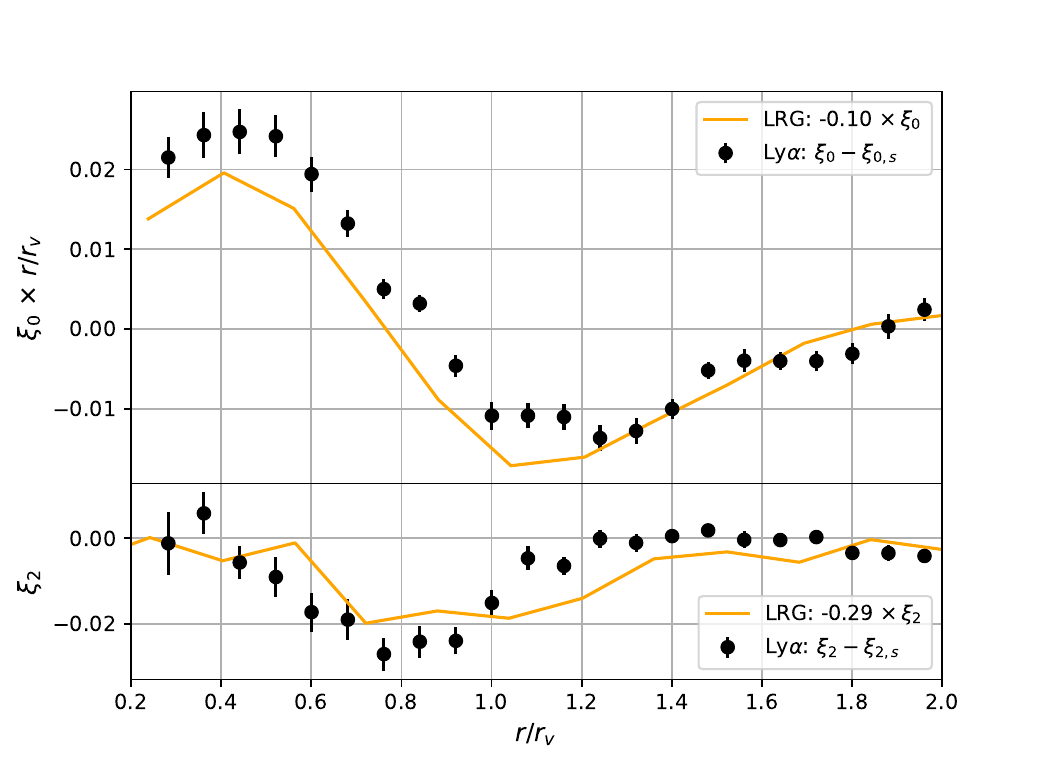}
    \caption{Points: measurements of the shuffling-corrected monopole and quadrupole functions for the void-forest correlation reported in Fig.~1 of the main article. Continuous lines: measurements of $\xi_0(r/r_v)$ and $\xi_2(r/r_v)$ for the LRG-void correlation, as reported from the SDSS DR16 sample in~\cite{Aubert2020}. We arbitrarily rescaled the LRG measurements along the $y$-axes by a factor of $-0.1$ (resp. $-0.29$) for $\xi_0$ (resp. $\xi_2$). The radial coordinate shown here is $r/r_v$ where we arbitrarily set $r_v=25\,\mpc$ in the case of the \lya-void correlation. }
\label{fig:cmp_galaxies}
\end{figure}

In order to highlight the similarity of our measurement with those derived from galaxy catalogs, we present here a qualitative but explicit comparison between the \lya-void correlation, and the galaxy-void correlation as derived from the SDSS DR16 LRG sample, using the measurement from~\cite{Aubert2020}. The comparison is shown in Fig.~\ref{fig:cmp_galaxies}.

As was explained in the main article and in Appendix B, for the \lya-void correlations we subtract the shuffled measurements in order to mitigate the impact of tomographic effects and noise.

The correlations measured in~\cite{Aubert2020} are expressed as a function of $r/r_v$, where $r_v$ is their void radii. In order to roughly match the correlation scales between both samples, in Fig.~\ref{fig:cmp_galaxies} we present our \lya-void correlation as a function of $r/r_v$, with $r_v$ arbitrarily set to $25\,\mpc$. This number is close to $r_v=22\,\mpc$, which we found in Appendix B when fitting the void profile formula of Eqn.~\ref{eq:voidprofile}. On the other hand, as explicated in the main article, the average radius of our void sample is $13\,\mpc$ in comoving units, as defined by our void finding algorithm. We attribute the difference by a factor of $\sim$ two to the different void radius definitions. In our case, this is set by the threshold $\delta_{F, {\rm tomo}} > 0.12$ applied to a map of the \lya~flux, while in the case of LRG it is set by the volume of Voronoi cells computed by the ZOBOV algorithm~\cite{Aubert2020}.

The monopole of the LRG-void correlation has a different sign, and differs by an order of magnitude compared to that of the \lya~forest. This is expected given the different biases of these two tracers, with the LRG bias $\sim 2$~\cite{Aubert2020} while that of the \lya~forest is $\sim -0.15$~\cite{duMasDesBourboux2020,Cuceu:2025nvl}. We choose to rescale the LRG-void correlation by a factor $s_0 = -0.10$ in order to roughly match the monopole amplitudes. After these two horizontal and vertical rescalings, we observe that the monopoles $\xi_0$ for both samples have the same qualitative patterns. We stress that this rescaling is not a fit: we expect the monopoles to differ anyway because the void finding algorithms are fundamentally different, so the details of the void profiles should differ especially on small scales.

To compare the quadrupoles in a way that is consistent with the monopoles, we rescaled the LRG-void quadrupole by a factor $s_2 = s_0 \times \frac{\beta_\alpha}{\beta_{\rm LRG}} = -0.29$, where $\beta_\alpha=1.21$ results from our measurement, and $\beta_{\rm LRG}=0.42$ was measured in~\cite{Aubert2020}. After this rescaling, the quadrupoles are of comparable amplitudes: this is expected because the RSD coefficients $\beta$ were obtained using very similar procedures in both cases. On the other hand, there is an apparent offset in the radial dependance of the quadrupolar pattern; we note that a similar offset is also seen when comparing our data with the \lya~mocks (see Fig.1 in the main article).

To summarize, we find that, after the shuffling correction, and taking into account the different radial scales and tracer biases, our measurement of the \lya-void correlation has similar patterns to that of SDSS LRGs.

\end{document}